\documentclass[a4,showpacs,amsmath,amssymb,preprint]{revtex4}

\usepackage{graphicx}%

\begin{document}
\title{Response of the Hodgkin-Huxley neuron
to a periodic sequence of biphasic pulses}

\author{L.~S.~Borkowski}
\affiliation{Department of Physics, Adam Mickiewicz University,
Umultowska 85, 61-614 Poznan, Poland}

\begin{abstract}
Electric stimulation of various parts of the nervous system
is a widely used therapeutic method and a principle of operation
of prosthetic devices. Its usefulness has been proven
in areas such as treatment of neurological disorders
and cochlear prostheses.
However the dynamic mechanisms underlying these applications
are not well understood. In order to shed some light
on this problem we study the response of the Hodgkin-Huxley (HH) neuron
subject to periodic train of biphasic rectangular current pulses.
One of the simpler ways to understand the behavior of a such a nonlinear
system is the analysis of the global bifurcation diagram
in the period-amplitude plane.
For short pulses the topology of this diagram
is approximately invariant with respect to the pulse polarity and shape details.
The lowest excitation threshold for charge-balanced input
was obtained for cathodic-first pulses
with an inter-phase gap (IPG) approximately equal $5 \textrm{ms}$.
The firing rate of the HH neuron stimulated at the frequency of its natural resonance
is a square root function of the pulse amplitude.
At nonresonant frequencies the quiescent state and the firing state coexist
and transition to firing is a discontinuous one.
We found a multimodal transition (MMT) in the regime
of irregular firing between the 2:1 and 3:1 locked-in states.
This transition separates the regime of odd-only multiples of the
stimulus period from the regime where modes of both parities
participate in the response.
A strong antiresonant effect was found between the states 3:1 and 4:1,
where the modes $(2+3n):1$, where $n=0,1,2,\ldots$, were entirely absent.
The antiresonant effects at high stimulation frequency,
such as MMT, may provide an explanation
for the therapeutic mechanism of deep brain stimulation.
\end{abstract}
\maketitle

\section{Introduction}
Stimulation of neural fibers with a train of electric current
pulses has become standard procedure in several areas
of clinical practice. This encompasses, but is not limited to,
cochlear and retinal prostheses \cite{SchmidtBak1996,Miller1998,Humayun1999},
deep brain stimulation (DBS) \cite{Benabid1991,McIntyre2000},
as well as high-frequency conduction block
of peripheral nerves \cite{TaideGroat2005}.
Typically such devices should minimize
charge and energy per pulse. The minimum current amplitude
required for the neuron to spike and its dependence
on stimulation frequency and pulse shape are among
the basic properties measured in experiment.
The injected charge should be kept at safe levels, mainly
to avoid changes in the chemical content of tissue near
the electrode which could damage tissue \cite{Yuen1981,McCreery1990} or cause
the electrode corrosion \cite{Merrill2005}.
Usually the injected charge needed by the neuron to spike
is minimized by the shortest possible width whereas optimization of the pulse energy
occurs at certain larger
width \cite{Wessale1992,Sahin2007,DanzlNabiMoehlis2010,Foutz2010,Jezernik2010}.
Selection of the most efficient waveform shape and duration,
such that delivered charge, energy and power are simultaneously optimized
is a difficult task \cite{Wongsarnpigoon2010b,Wongsarnpigoon2010a}.
Most studies indicate that non-rectangular pulses are more energy efficient
than rectangular ones \cite{Sahin2007,Foutz2010,Jezernik2010,Wongsarnpigoon2010a}.

The general stimulation waveform in DBS applications consists
of a brief cathodic pulse, followed by a delay, and then
a charge-balancing anodic pulse \cite{Butson2007}.
The change of the waveform affects stimulation thresholds of local cells and fibers
of passage differently, thus aiding in selecting the stimulation target.
Both symmetric and asymmetric biphasic charge-balanced
waveforms in selective stimulation of the CNS were investigated \cite{McIntyre2000,McIntyre2002}.
Also most cochlear implants use charge-balanced biphasic (BP) stimulation.
Studies of auditory nerve fibers (ANF)
found that threshold can be reduced \cite{Shepherd1999,Carlyon2005}
by introducing a delay between standard BP pulses
commonly used in cochlear implants.

Some of the DBS research is focused on the dynamics
of the cortico-basal-ganglia-thalamo-cortical
network. It is thought that optimal response to DBS occurs at resonant frequencies
of the network \cite{Montgomery2008,Montgomery2007,Grill2004}.
However, DBS is less effective at lower
frequencies and the therapeutic effect is most beneficial above $100 \textrm{Hz}$,
well above the frequency range normally considered for basal ganglia processing.
Since there is some evidence of $300 \textrm{Hz}$ subthalamic oscillations \cite{Foffani2003}
it has been suggested that multiple natural periods may be involved in parkinsonism
due to multiple network loops.
However, there is another possible mechanism for disruption of network activity patterns.
Recent studies of resonant neurons revealed the presence of MMT,
occuring at approximately $2.5f_{res}$, where $f_{res}$ is the natural resonance frequency,
where the parity of response modes is changed \cite{Borkowski2009,Borkowski2010,Borkowski2011,Borkowski2012}.
This effect has an antiresonant character.
It involves both a chaotic behavior and a significant slowing down of average response frequencies
in the vicinity of MMT. 
It would therefore be useful to investigate if this phenomenon is sensitive
to details of the stimulus waveform and whether it exists also for charge-balanced inputs.

Studies of electrically stimulated ANFs show that at low stimulation rates
these fibers fire regularly, in-step to applied stimulus, whereas at high stimulation rates
the response is highly irregular.
Irregular firing and desynchronization among a group of ANFs is usually attributed to irregular
synaptic input and physiological noise \cite{Morse1996,Moss1996}.
It must be kept in mind however that neurons themselves are strongly nonlinear systems
and are capable of irregular firing even
in the absence of noise \cite{Borkowski2009,Borkowski2010}.
O'Gorman et al. showed recently \cite{OGorman2009,OGorman2010}
that a dynamic instability is a plausible explanation
of firing irregularities in stimulated ANF.
Using the FitzHugh-Nagumo (FHN) model \cite{FitzHugh1961},
they obtained a positive Lyapunov exponent which is consistent with experiments, where
mutual desynchronization between similarly stimulated fibers was observed.
Their study is consistent with our analysis of the MMT within the HH
\cite{Borkowski2009,Borkowski2010,Borkowski2011}
and Morris-Lecar \cite{Borkowski2012} models. 
Since multimodal dynamics is common in electrically stimulated ANFs
\cite{Rose1967,Moxon1971,Hochmair1984,vandenHonert1987,Litvak2003a,Litvak2003b,Miller2008},
the current article may help understand
the response of ANFs to stimulation by a train
of high-frequency current pulses.

The aim of this work is not the search for the most optimal
charge and energy-efficient pulse shape. The focus is on
(i) proving the invariance of the global bifurcation diagram of a resonant
neuron with respect to pulse shape, polarity, and IPG,
(ii) finding the optimal IPG for a rectangular biphasic waveform,
and (iii) proving that the dynamic instability in the form of the recently
discovered multimodal transition exists also for biphasic
pulses, which makes this phenomenon interesting in the context of clinical applications.

\section{Methods}
We analyzed the response of the HH neuron to periodic
biphasic charge balanced pulses.
The requirement of charge neutrality was imposed to avoid tissue damage.
Such stimuli were investigated
experimentally \cite{McKay2003,Carlyon2005,PradoGuitierrez2006}
and theoretically by several groups \cite{Shepherd1999,Carlyon2005,Macherey2007,Smit2010}.

We considered the model with the classic parameter set
and rate constants \cite{HH1952},
\begin{equation}
\label{HHeqn}
C\frac{dV}{dt}=-I_{Na}-I_K-I_L+I_{app},
\end{equation}
where $I_{Na}$, $I_K$, $I_L$, $I_{app}$,
are the sodium, potassium,
leak, and external current, respectively.
$C=1\mu\textrm{F/cm}^2$ is the membrane capacitance.
The input current was a periodic set
of rectangular steps of height $I_0$ and width $\tau$.
The calculations were carried out with the time step of $0.001 \textrm{ms}$
and were run for $40 \textrm{s}$, discarding the initial $4 \textrm{s}$.
The differential equations were integrated using
the fourth-order Runge-Kutta scheme implemented by the author.
None of the publicly available programs were used.

\begin{figure}[ht]
\includegraphics[width=0.35\textwidth]{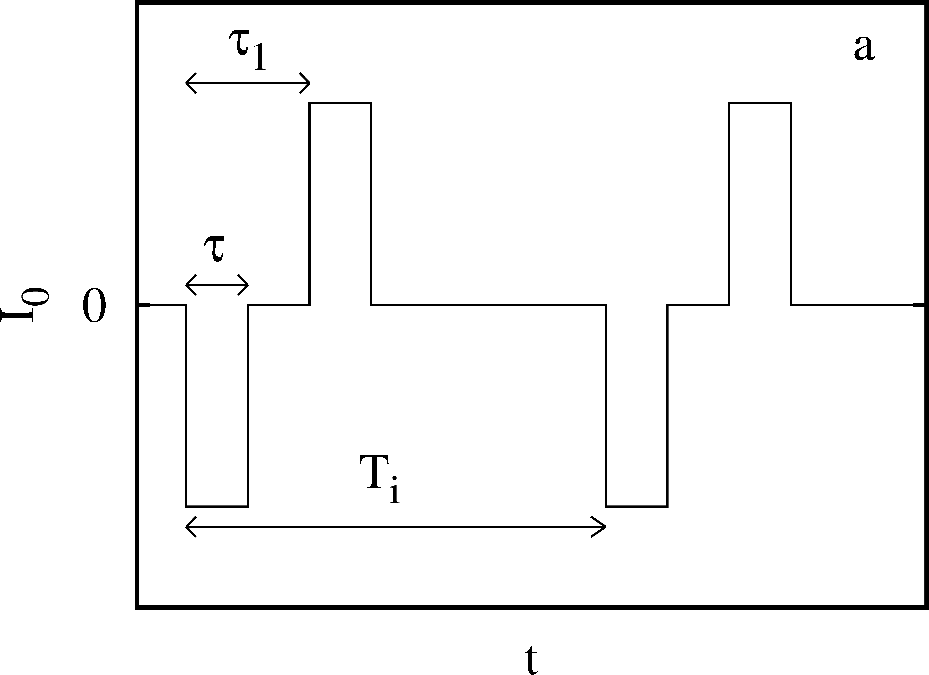}
\vskip 0.5cm
\includegraphics[width=0.35\textwidth]{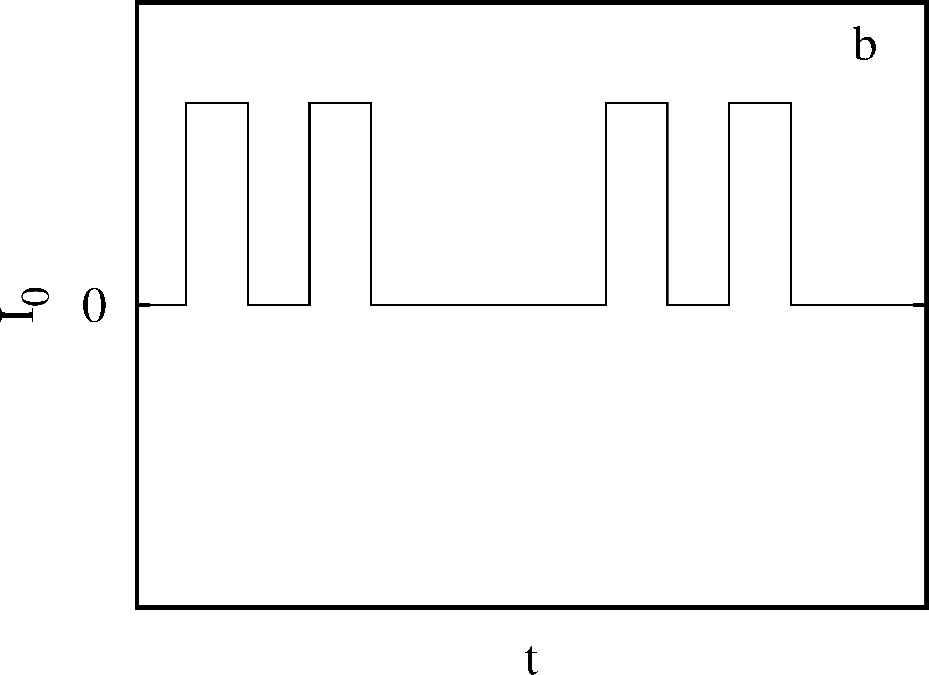}
\caption{Stimulus waveform used in the calculations: (a) charge-balanced
biphasic form with delay, and (b) monophasic pulses with delay.
The delay between the onset of the cathodic and anodic phase is $\tau_1$
and $\tau$ is the width of each phase.
}
\label{stimulus}
\end{figure}

Fig. \ref{stimulus} shows the form of biphasic pulses
used in the calculation. The emphasis of this work is on the charge-balanced
stimulus shown in Fig. \ref{stimulus}a.
We investigated how the amplitude,
period $T_i$, width $\tau$
and delay $\tau_1$ between the onsets of the two phases,
influence the neuron's response.
Stimuli of this form are used
in cochlear \cite{Shepherd1999,Carlyon2005,Macherey2007,Smit2010}
and visual \cite{SchmidtBak1996,Humayun1999} neural prostheses,
as well as deep brain stimulation \cite{Foutz2010}
and muscle stimulation \cite{Merrill2005}.
Electrical stimulation protocols have been used to learn
about cortical function \cite{Tehovnik2006} and vestibulo-ocular reflex
eye movements in chinchillas \cite{Davidovics2011}.
High-frequency biphasic current pulses may also be useful
in reversible peripheral nerve block that would
be desirable in some clinical applications \cite{TaideGroat2005}.
We also considered monopolar biphasic
pulses, shown in Fig. \ref{stimulus}b,
and calculated threshold dependence on IPG at different
stimulation frequencies.

\section{Results}
\begin{figure}[ht]
\centering
\includegraphics[width=0.45\textwidth]{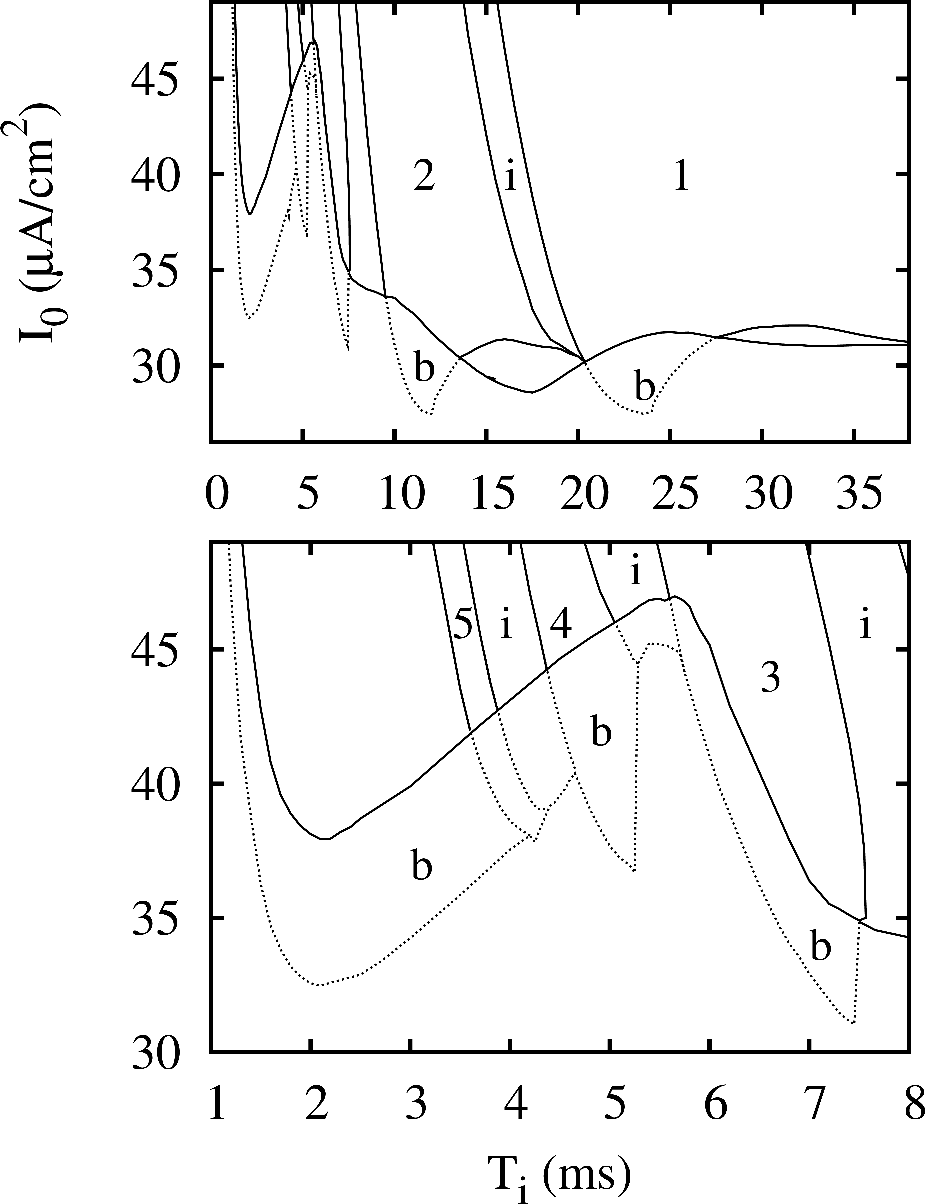}
\caption{Top panel: Response diagram for a biphasic stimulus with no delay
between the anodic and the cathodic part of a pulse,
$\tau=\tau_1=0.6\textrm{ms}$ The high frequency limit of the same diagram
is shown in more detail in the bottom panel.
The numbers $1,2,\ldots$ indicate the zones of 1:1, 2:1, and higher
locked-in states; symbols $i$ and $b$ are used to denote regions
of irregular and bistable behavior, respectively. 
Borders of bistable regions are marked with a dotted line.
In the zones between the solid and dotted lines the firing state
coexists with the silent one.
The resonance near $T_i = 17\textrm{ms}$ and $I_0=30\mu\textrm{A/cm}^2$, where
the minimum of the excitation threshold occurs, is dominated
by the 3:1 state and higher order states. Its mode-locking structure is essentially
identical to that of monophasic monopolar stimulation shown in Fig. 3 of Ref. \cite{Borkowski2011}.
The firing rate $f_o/f_i$, where $f_o$ and $f_i$ are
the average response and stimulation frequency, 
continuously decreases to 0 at the tip of the resonance, $f_o/f_i \sim (I_0-I_{th})^{1/2}$.
The second resonance near $T_i = 34\textrm{ms}$ is 
occupied by states of order higher than 1:1 (see also Ref. \cite{Borkowski2011}).
Also at this resonance the firing rate depends continuously on $I_0$.
Bottom panel: The regime of high stimulation frequency in more detail.
There is a strong antiresonance near $T_i=5.5\textrm{ms}$.
}
\label{fig2}
\end{figure}

Fig. \ref{fig2} shows the global bifurcation diagram
in the period-amplitude plane
for a charge-balanced biphasic current with no delay between
the cathodic and anodic part, i.e. $\tau=\tau_1=0.6\textrm{ms}$. The numbers in the figure
show the location of various states locked-in to the stimulus. For example 2
denotes the state 2:1, where there is one action potential for each two
stimulus pulses. The notation p:q means q output spikes for every p current pulses.
The type of the near-threshold response depends strongly on stimulation frequency. 
The firing rate $f_o/f_i$ is a continuous function of the current amplitude near
the resonance \cite{Borkowski2011}, $T_i\sim T_{res}\simeq 17\textrm{ms}$,
where $T_{res}$ is the neuron's preferred interspike interval (ISI).

Resonant stimulation periods are those at which the membrane voltage
attains its maxima \cite{Puil1986,Hutcheon2000}. In the context of this article we prefer
to use a related but slightly different definition.
The resonances can be identified by the minima of the excitation threshold.
In Fig. \ref{fig2} they are located
at $T_i \simeq 17 \textrm{ms}$ and $34 \textrm{ms}$. The local threshold maxima
are associated with antiresonances. In Fig. \ref{fig2}
they occur at $T_i \simeq 25 \textrm{ms}$
and just below $6 \textrm{ms}$.
It is clear from Fig. \ref{fig2} and from earlier
analysis \cite{Borkowski2011,Borkowski2012} that antiresonances are
accompanied by bistable behavior. Most of the regime of high stimulation
frequency has antiresonant character.

In bistable regions the neuron can be in one of two states.
One is a firing state, such as 1:1, 2:1, 3:1, etc.
and the other is the quiescent state.
Boundaries of these areas were found using a continuation algorithm,
in which end values from previous iteration
were used as initial conditions for a new parameter set.

\begin{figure}[ht]
\includegraphics[width=0.45\textwidth]{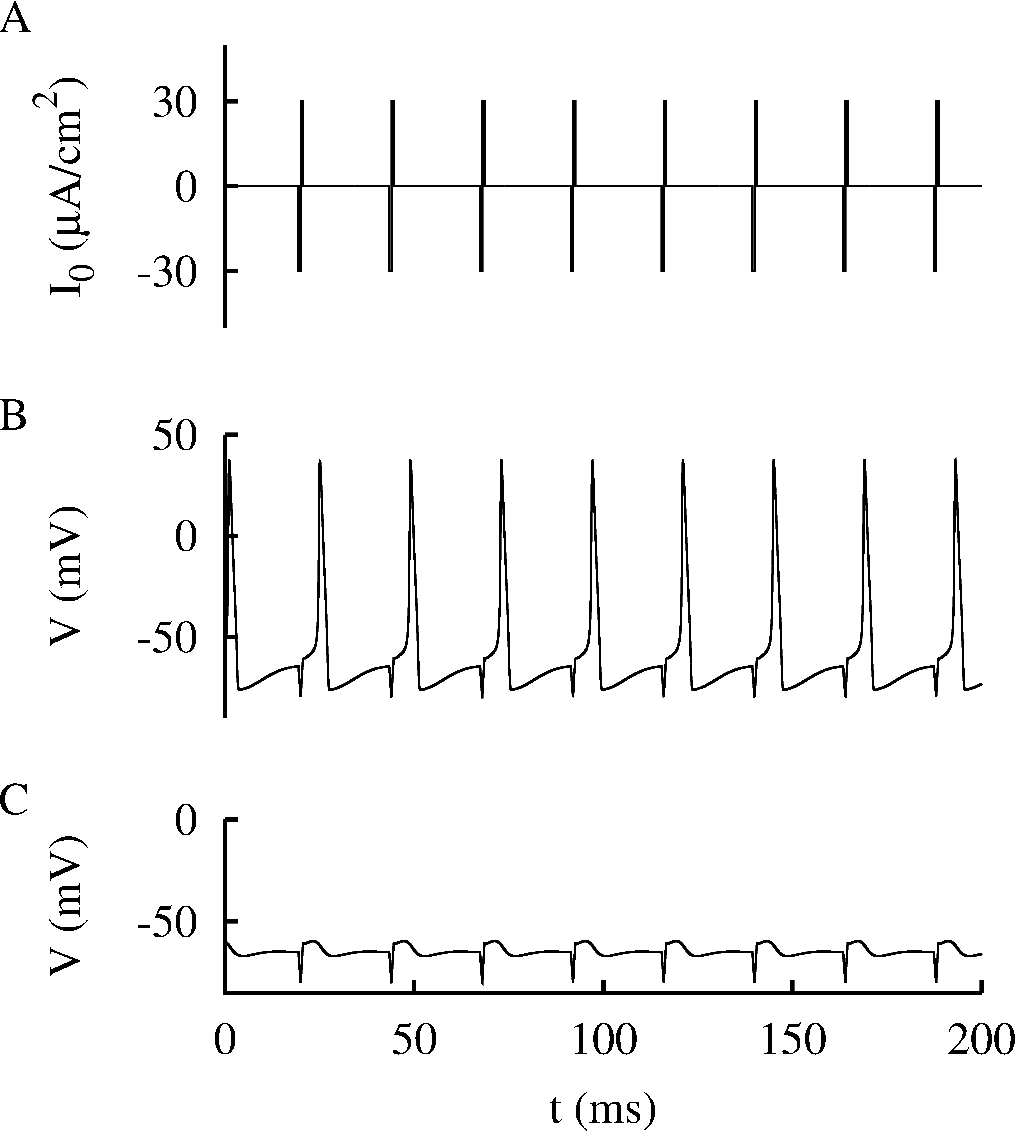}
\includegraphics[width=0.45\textwidth]{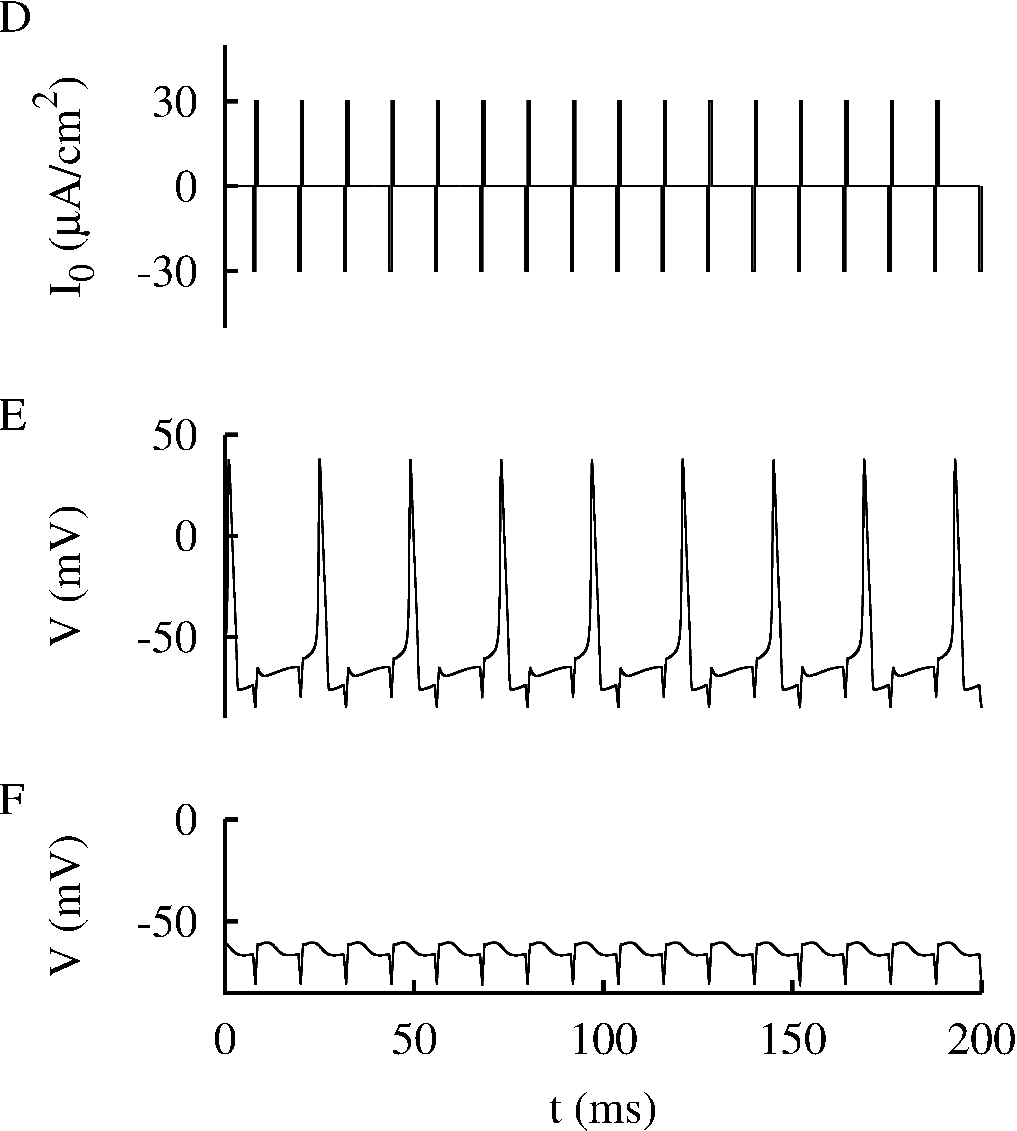}
\caption{Examples of solutions in bistable zones for stimulus with no inter-phase gap,
$\tau=\tau_1=0.6\textrm{ms}$, and amplitude $I_0=30\mu\textrm{A/cm}^2$,
and stimulus period \textbf{A} $T_i=24\textrm{ms}$, \textbf{D} $T_i=12\textrm{ms}$.
\textbf{B} Firing solution locked 1:1 in response to stimulus shown in \textbf{A}.
\textbf{C} Non-firing solution in response to stimulus \textbf{A}.
\textbf{E} Firing solution locked 2:1 in response to stimulus shown in \textbf{D}.
\textbf{F} Non-firing solution in response to stimulus \textbf{D}.
}
\label{fig3}
\end{figure}

Periodically stimulated neural oscillators are known to phase-lock with
the stimulus. Examples of such locking are shown
in Fig. \ref{fig3}.
The 1:1 locking is displayed in Fig. \ref{fig3}a,
where there is an action potential for every stimulus pulse.
Mode-locking of order $n$:1, where $n$ is an integer greater than 1,
can be obtained for higher stimulation frequencies
or just above threshold for $T_i\simeq T_{res}$ \cite{Borkowski2011}.
Fig. \ref{fig3} shows
an example of 2:1 locking for $T_i=12\textrm{ms}$.
Both regular and irregular patterns, corresponding
to noninteger ratios of the average output period $T_o$
to the input period $T_i$, also exist.

The overall topology of this diagram closely
resembles the result for monophasic monopolar pulses \cite{Borkowski2011}.
In the high frequency limit the threshold is a nonmonotonic
function of $T_i$ with a local maximum near $T_i=5.5\textrm{ms}$,
which is approximately $1/3$ of the resonance period $T_{res}$.
In other words, the HH neuron is least likely to respond
when driven with the frequency $3f_{res}$.
Below $T_i=8\textrm{ms}$ the entire perithreshold
region is bistable. Here the quiescent state coexists
with either a locked-in or a chaotically firing state.
Which of these solutions is obtained depends
sensitively on initial conditions.
It is worth noting that nonmonotonic behavior of the hearing threshold
was observed in experiments
in cochlear implant users \cite{vanWieringen2006}.
However more detailed experimental studies of threshold dependence on stimulation frequency
are needed before comparisons could be made. Special attention must be paid
to the possibility of bistable behavior at the threshold.

\begin{figure}[ht]
\center
\includegraphics[width=0.45\textwidth]{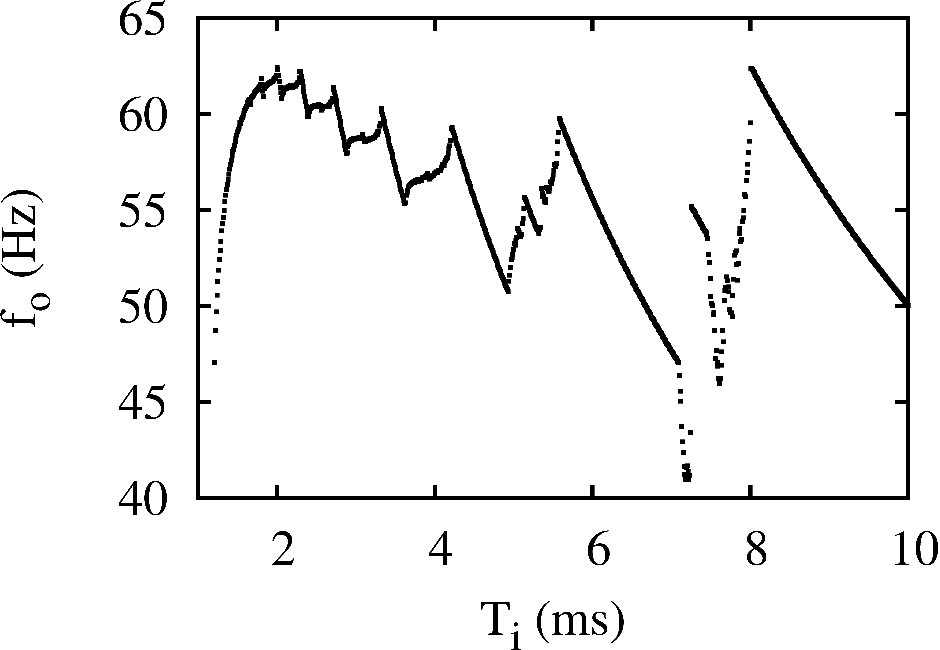}
\caption{The firing frequency in the limit of high stimulation
frequency for a biphasic pulse at $I_0=47\mu\textrm{A/cm}^2$, with no
inter-phase delay, $\tau = \tau_1 = 0.6\textrm{ms}$.
The response between the mode-locked intervals is nonmonotonic
and highly irregular.
}
\label{fig4}
\end{figure}

\begin{figure}[ht]
\includegraphics[width=0.7\textwidth]{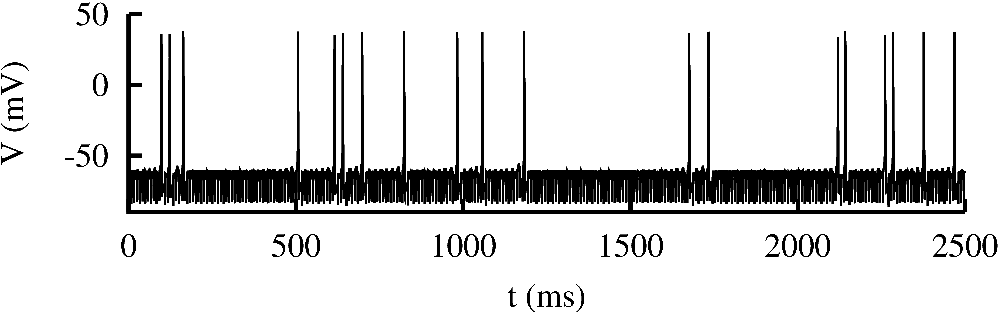}
\caption{Sample $V(t)$ run for $T_i = 8.35\textrm{ms}$, very close
to the multimodal transition.
Here $\tau=\tau_1=0.6\textrm{ms}$, and the stimulus amplitude is
$I_0=35\mu\textrm{A/cm}^2$.
The shown ISI sequence is $3,5,41,13,3,7,15,19,9,15,59,7,46,3,14,3,11,11$, in units of $T_i$.
The ISIs are mostly odd multiples of $T_i$.
}
\label{fig5}
\end{figure}

The firing frequency depends nonmonotonically on $T_i$.
Fig. \ref{fig4} shows $f_o$ vs. $T_i$
for $I_0$ slightly above threshold. The largest deviation
from the resonant frequency occurs between the $\textrm{2:1}$
and $\textrm{3:1}$ states, near $T_i=7.5\textrm{ms}$.
This is a signature of a dynamic
instability associated with the competition
of even and odd modes \cite{Borkowski2009}.
Sample $V(t)$ run very close to this instability,
is shown in Fig. \ref{fig5}.
The odd multiples of the stimulus period clearly dominate. Out of 18 ISIs visible
in Fig. \ref{fig5} there are
only two even multiples of $T_i$ in Fig. \ref{fig5}.

\begin{figure}[ht]
\center
\includegraphics[width=0.45\textwidth]{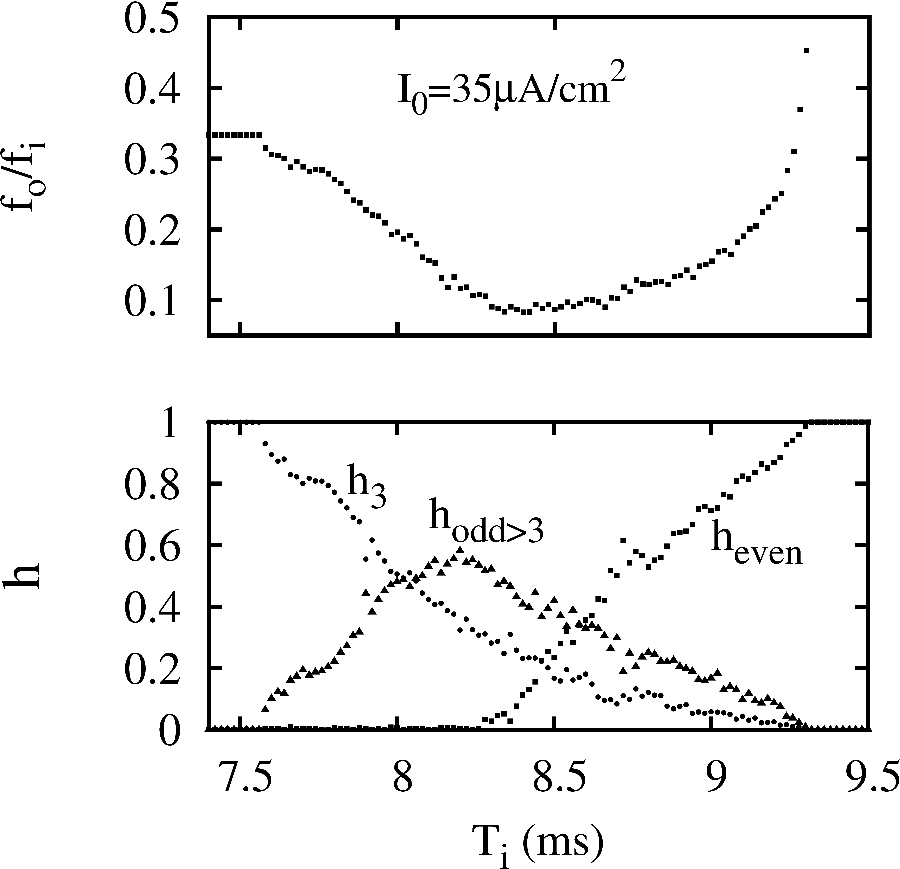}
\caption{The firing rate (top) and the histogram weight
of odd and even modes (bottom) in the vicinity
of the odd-all multimodal transition, $T_i \simeq T_{mm}$,
for stimulation with a sequence of biphasic pulses
of amplitude $I_0=35\mu\textrm{A/cm}^2$.
Even modes appear only above the minimum of $f_o/f_i$.
An interspike interval is classified as belonging
to the ${n}^{th}$ mode if it falls between
$(n-1/2)T_i$ and $(n+1/2)T_i$.
}
\label{fig6}
\end{figure}

In an earlier article we showed
that the HH neuron responding to a periodic sequence
of monophasic pulses undergoes a multimodal
odd-all \cite{Borkowski2009}
and even-all transition \cite{Borkowski2011}
in some regions of parameter space.
Fig. \ref{fig6} shows the minimum of the firing rate
occuring between the 2:1 and 3:1 locked states
for pulse amplitude $I_0=35\mu\textrm{A/cm}^2$.
The weight of even modes
is nonzero only for $T_i$ above the minimum of $f_o/f_i$.

For $T_i$ below $T_{mm}$ only odd modes exist.
The role of the mode parity is easier to understand if we remember
that $2T_i$, where $T_i\simeq 8.5 \textrm{ms}$
is approximately equal to the resonant period $T_{res}$. The competition
of odd and even modes in Fig. \ref{fig6}
accompanies the transition between the resonant
and antiresonant regime. The vanishing of even modes
below $T_i\simeq 8.4\textrm{ms}$ and a significant decrease
of the firing rate are a clear signature of entering the antiresonant regime.

\begin{figure}[ht]
\center
\includegraphics[width=0.45\textwidth]{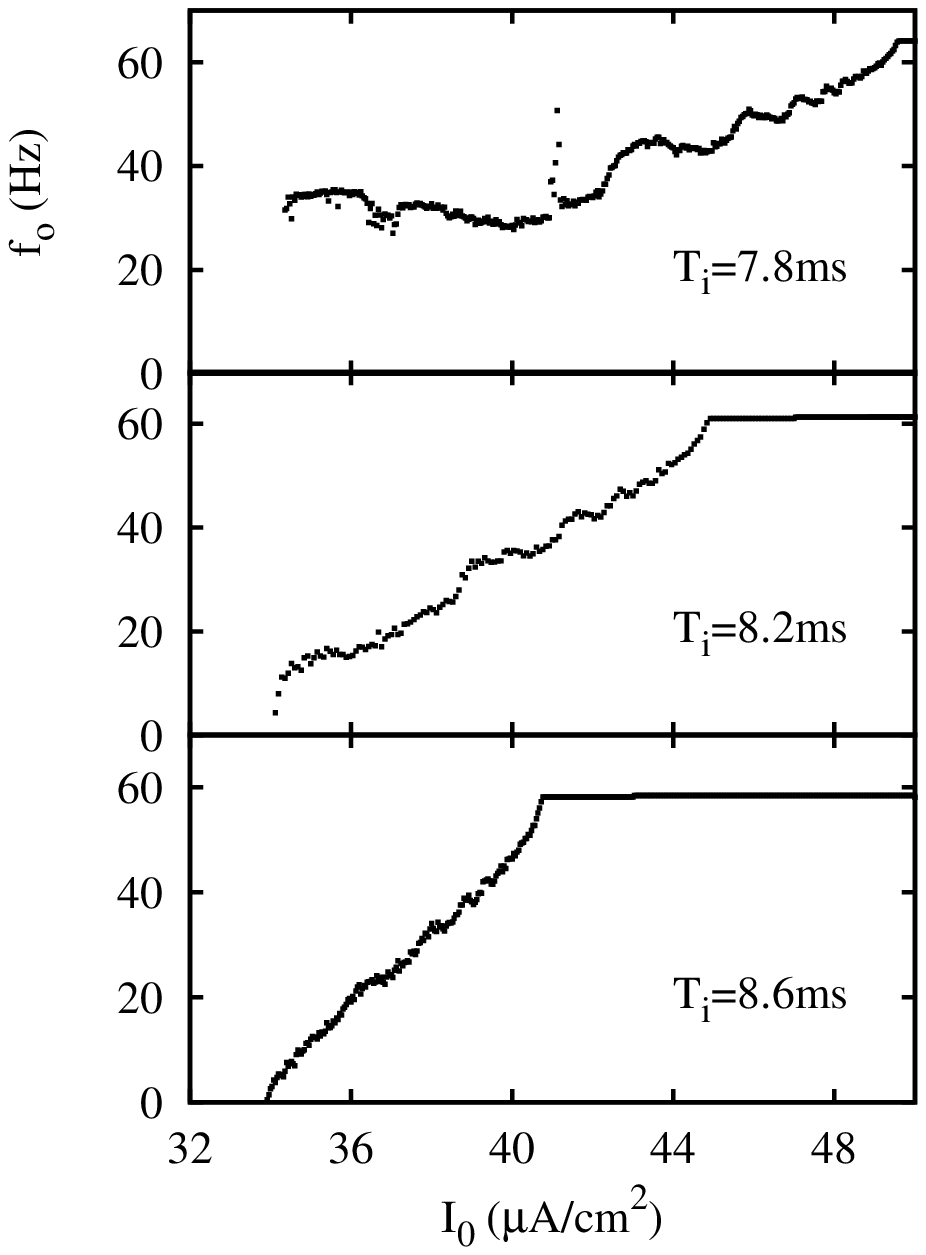}
\caption{The firing rate as a function of the stimulus amplitude $I_0$ at fixed
$T_i$ near the multimodal transition. The transition occurs at $T_{mm} \simeq 8.4\textrm{ms}$.
Here $\tau=\tau_1=0.6\textrm{ms}$.
}
\label{fig7}
\end{figure}

The dependence of the firing rate on
pulse amplitude $I_0$ in the vicinity of the threshold is qualitatively different on both sides
of the multimodal transition, see Fig. \ref{fig7}.
For $T_i<T_{mm}$ the firing rate depends discontinuously
on pulse amplitude. However the size of the discontinuity decreases as $T_i$ approaches
$T_{mm}$. For $T_i \ge T_{mm}$, $f_o$ is a continuous, nearly linear function of $I_0$.
In this regime the average firing rate may be arbitrarily small.

\begin{figure}[ht]
\center
\includegraphics[width=0.45\textwidth]{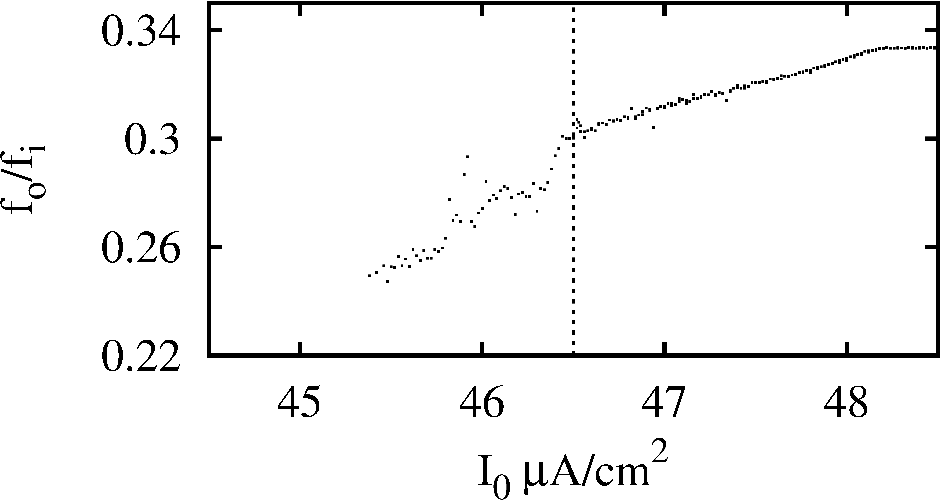}
\caption{The firing rate as a function of the stimulus amplitude for $T_i=5.5\textrm{ms}$
and no delay between the phases of the pulse, $\tau=\tau_1=0.6\textrm{ms}$.
The vertical broken line separates bistable and monostable regions.
}
\label{fig8}
\end{figure}

At high frequencies, for $T_i < 5.5 \textrm{ms}$,
the bistable zones encompass not only phase-locked
states with integer ratios of $T_o/T_i$, but also states
with no regular response pattern. Dependence
of the firing rate the current pulse amplitude in one of such zones
is shown in Fig. \ref{fig8}. Here the irregularly
firing state coexists with a quiescent state.
The firing rate depends discontinuously on stimulus
amplitude in this regime.

\begin{figure}[ht]
\includegraphics[width=0.45\textwidth]{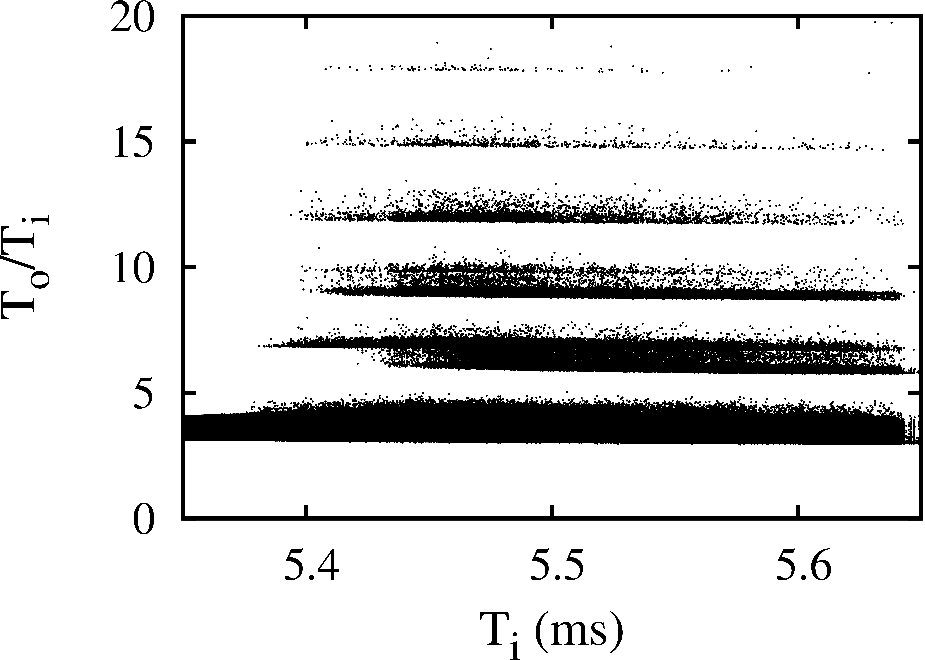}
\hskip 1cm
\includegraphics[width=0.45\textwidth]{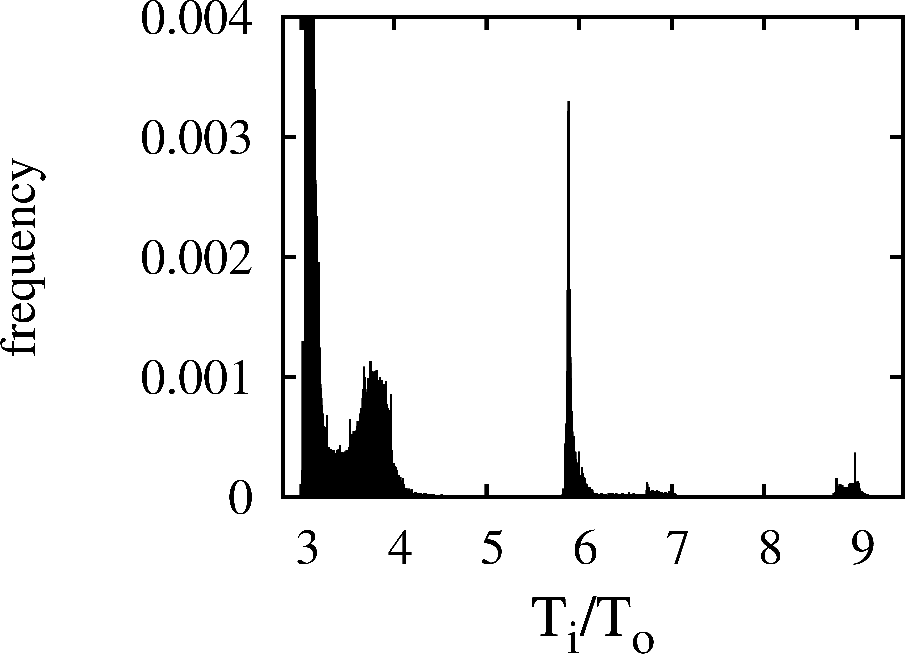}
\caption{Left:
The ratio of the output period $T_o$ to the input period
$T_i$ between the 3:1 and 4:1 entrained states
for a biphasic pulse at $I_0=45.8\mu\textrm{A/cm}^2$.
Right: Histogram of response modes at $T_i=5.55\textrm{ms}$
and $I_0=45.8\mu\textrm{A/cm}^2$.
The calculations were carried out for stimulus
with no inter-phase gap, $\tau=\tau_1=0.6\textrm{ms}$.
The histogram bin size is $0.01\textrm{ms}$.
Note the absence of modes 5,8,11,14, and 17.
}
\label{fig9}
\end{figure}

\begin{figure}[ht]
\center
\includegraphics[width=0.45\textwidth]{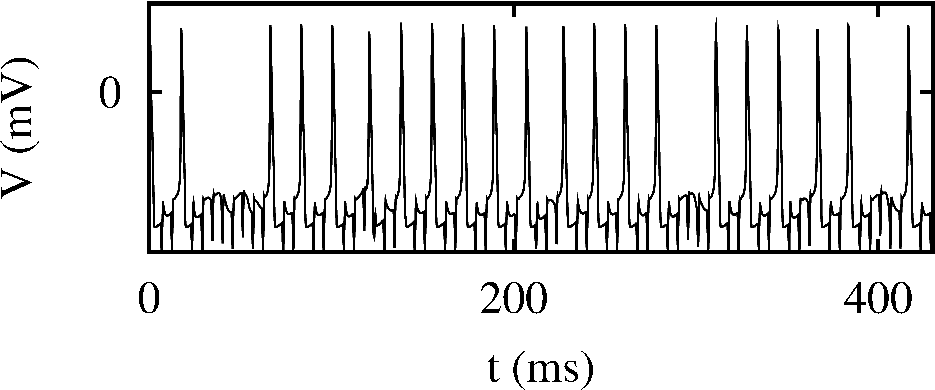}
\caption{Sample $V(t)$ dependence for $T_i = 5.55\textrm{ms}$.
Here $\tau=\tau_1=0.6\textrm{ms}$, and the stimulus amplitude 
$I_0=45.8\mu\textrm{A/cm}^2$ is slightly above threshold.
The modes 3:1 and 4:1 dominate, with no clear separation between them.
The modes $(5+3n):1$, where $n=0,1,2,\ldots$, do not appear in the neuron's response.
}
\label{fig10}
\end{figure}

Another interesting effect occurs between the 3:1 and 4:1
states at high frequencies. Bands of $T_o$ vs. $T_i$ values
are shown in Fig. \ref{fig9}.
Note the absence of all modes of the form $(5+3n):1$, where
$n$ is a nonnegative integer. This is a consequence
of the fact that the product $(5+3n)T_i$ falls
in an antiresonant regime for $T_i\simeq 5.5\textrm{ms}$.
A typical $V(t)$ dependence in this situation is shown in
Fig. \ref{fig10}.
Since this is a qualitative effect, independent of a particular
pulse shape, it may provide
a strong test of applicability of the Hodgkin-Huxley type dynamics
in various contexts.

\begin{figure}[ht]
\center
\includegraphics[width=0.45\textwidth]{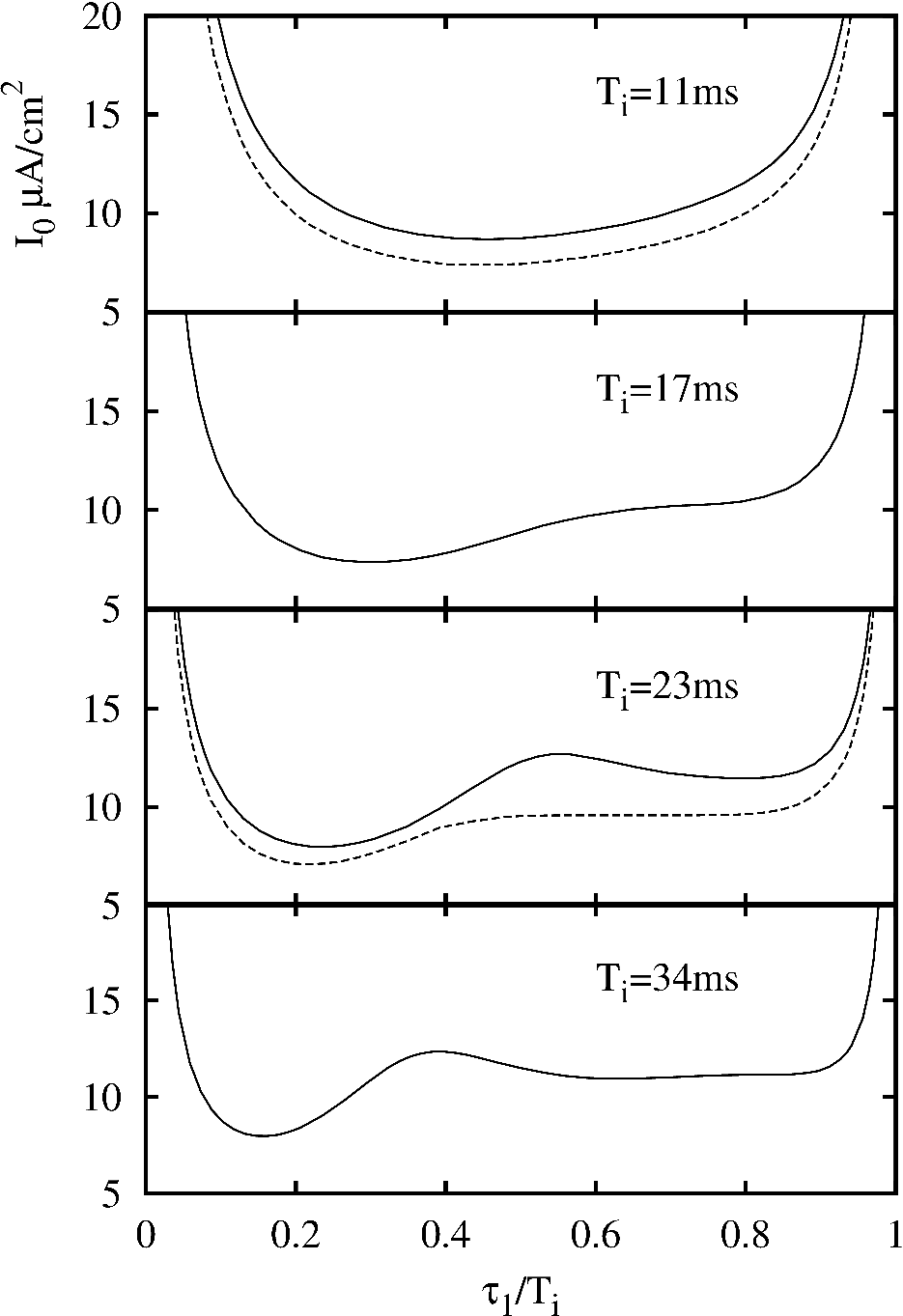}
\caption{Excitation edge for different interpulse separation
as a function of time difference between the cathodic and anodic
phase. From top to bottom:
$T_i = 11, 17, 23,$ and $34 \textrm{ms}$,
for stimulation by rectangular current pulses
of width $\tau = 0.6 \textrm{ms}$ and height $I_0$.
For different pulse frequencies the minimum occurs
at $\tau_1 \simeq 5 \textrm{ms}$.
The threshold is bistable for $T_i=11\textrm{ms}$
and $23\textrm{ms}$, where both the firing solution and the quiescent state
exist between the broken line and the solid line.
}
\label{fig11}
\end{figure}

Fig. \ref{fig11} shows the excitation threshold as a function
of the delay between the anodic and the cathodic phases of the pulse,
for four choices of the stimulus period.
In each case the optimum $\tau_1$ is about $5 \textrm{ms}$.
If the input period is in the antiresonant regime,
i.e. for $T_i=11$ and $23\textrm{ms}$,
there are two thresholds because of bistability,
both of which have minima at similar values of $\tau_1$.
The upper threshold where the transition to firing
occurs through a subcritical Hopf bifurcation,
has a maximum at approximately $\tau_1=13\textrm{ms}$.
This antiresonant feature occurs
at the same $\tau_1$ also in stimulation by pulses separated by $T_i=34\textrm{ms}$.
Lower thresholds for biphasic cathodic-first stimuli were also
observed in Refs. \cite{vanWieringen2005,vanWieringen2008,Macherey2006}
and calculated by Smit et al. \cite{Smit2010} in an ANF model
containing persistent sodium and slow potassium currents.

\begin{figure}[ht]
\center
\includegraphics[width=0.45\textwidth]{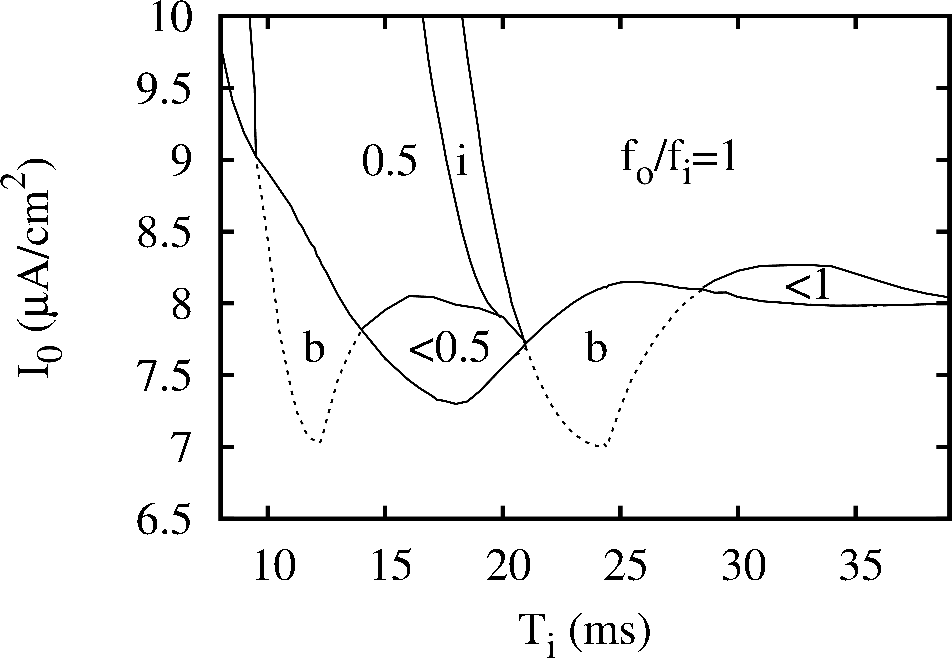}
\caption{Response diagram for stimulus with an optimal inter-phase gap,
$\tau_1=5\textrm{ms}$,$\tau=0.6\textrm{ms}$. Symbols $i$ and $b$
indicate a zone of irregular firing and bistability, respectively.
Boundaries of bistable zones are marked with a dotted line.
The resonance area near $T_i = 17\textrm{ms}$ is dominated
by the 3:1 state and higher order states. In this zone
the firing rate is below 0.5 and decreases almost continously to 0
at the tip of the resonance.
The second resonance near $T_i = 34\textrm{ms}$ is 
occupied by modes higher than 1:1. In this regime the firing rate
is below 1.
}
\label{fig12}
\end{figure}

The minimum threshold is obtained for the stimulus frequency
tuned to the natural resonance of the neuron.
The resonance period sets also an upper limit
of IPG, for which the interaction between
the cathodic and anodic phase is noticeable, see bottom panel
of Fig. \ref{fig11}.
The spike-triggered average of stimulus $I(t)$ for isolated
spikes in the HH model driven by Gaussian random noise current
with a short correlation time also indicated
the preference of the neuron for approximately $5\textrm{ms}$
separation between the negative and positive
parts of the current \cite{Arcas2003}.
This is a reflection of the HH neuron's internal
dynamics in which the sodium channel remain
open for about $5\textrm{ms}$ from the stimulus onset.
The bifurcation diagram with an optimal IPG,
$\tau_1\simeq 5\textrm{ms}$, shown in Fig. \ref{fig12}
closely resembles Fig. \ref{fig2}. The only significant difference
is the shift of the boundary between the quiescent state
and firing states to lower values of $I_0$.

\begin{figure}[ht]
\center
\includegraphics[width=0.45\textwidth]{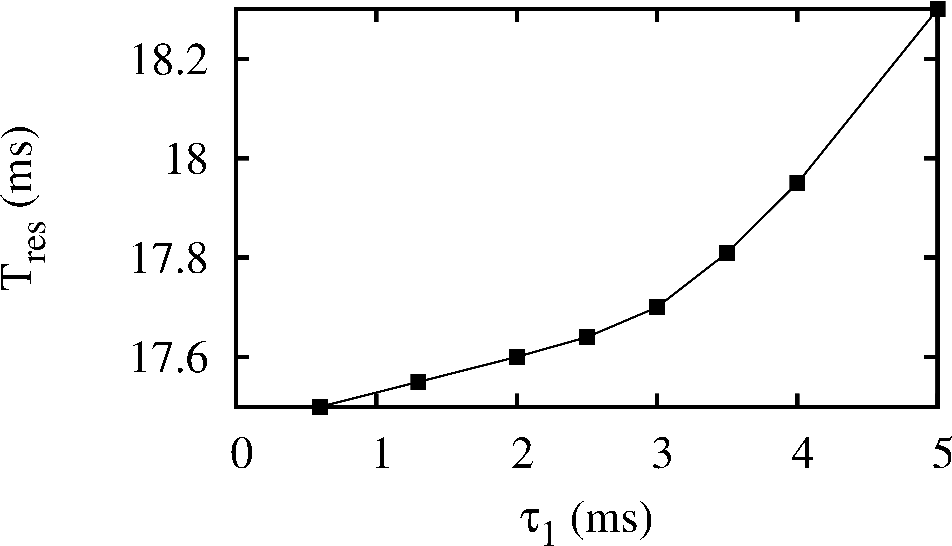}
\caption{Dependence of the location of the resonance tip on the inter-phase gap $\tau_1$.
The phase width is fixed at $\tau=0.6\textrm{ms}$.
}
\label{fig13}
\end{figure}

The precise location of the tip of main resonance $T_{res}$ depends on the size of IPG.
For fixed $\tau$ the resonance period grows with increasing $\tau_1$, see
Fig. \ref{fig13}.
This slowing down is due the cathodic pulse
which delays the firing cycle.
At the optimal IPG, $\tau_1=5\textrm{ms}$,
the tip of the resonance is shifted
by approximately $1\textrm{ms}$ relative to its position
for monophasic excitatory pulses.

\begin{figure}[ht]
\center
\includegraphics[width=0.45\textwidth]{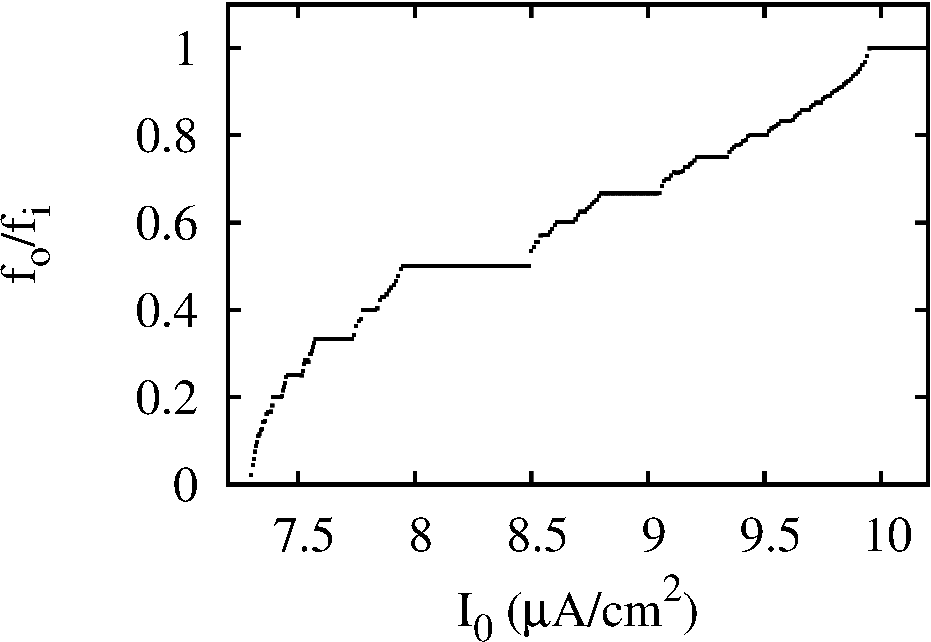}
\caption{Firing rate as a function of the stimulus amplitude
in the resonant regime, $T_i=18.3 \textrm{ms}$.
Here $\tau=0.6\textrm{ms}$, $\tau_1=5\textrm{ms}$.
Note the continuous dependence of the firing rate on $I_0$,
$f_o \sim (I_0-I_{th})^{1/2}$, near the threshold.
}
\label{fig14}
\end{figure}

The firing rate at the main resonance frequency, shown in Fig. \ref{fig14},
is a square root function of the pulse amplitude
$f_o \sim (I_0-I_{th})^{1/2}$, where $I_{th}$ is an approximate
threshold. This implicates that the transition to firing
occurs via a saddle node bifurcation. This is a property of the neuron's
internal dynamics and is independent of a stimulus shape.
However, it is important to realize that the character of the bifurcation
at the resonance changes as the width of the pulse increases beyond
optimal\cite{Borkowski2011}. For pulse widths approaching the time scale
of the resonance, the firing rate becomes a discontinuous function
of $I_0$. The transition to firing occurs then via a subcritical Hopf bifurcation
and there is a bistable zone near the threshold at all frequencies.

\begin{figure}[ht]
\center
\includegraphics[width=0.45\textwidth]{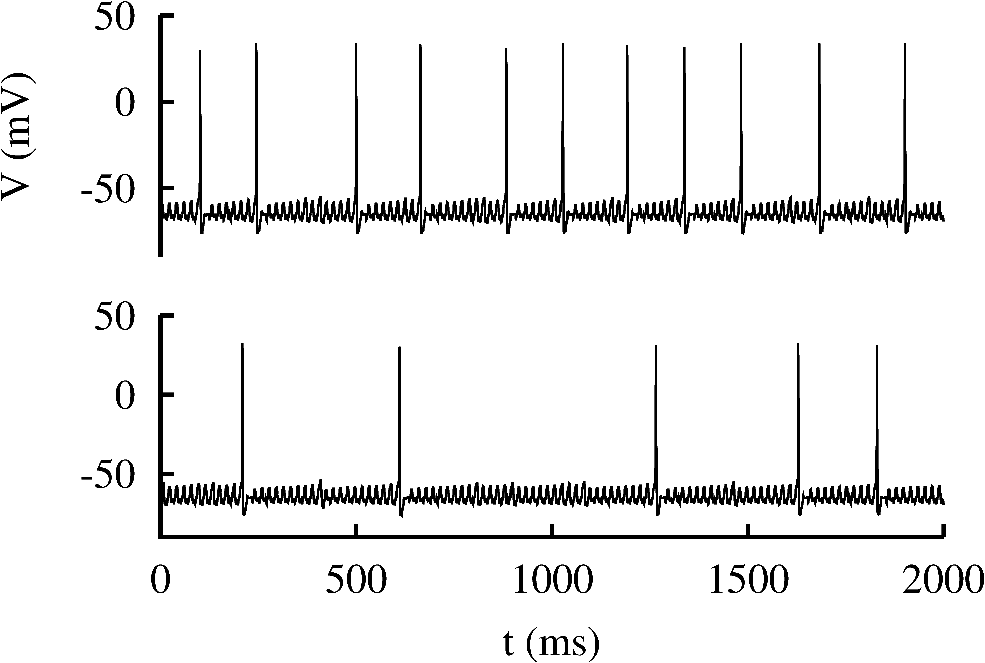}
\caption{Sample $V(t)$ dependence near the resonance at
$T_i=18.2 \textrm{ms}$, for stimulation by biphasic pulses with $\tau=0.6\textrm{ms}$, $\tau_1=5\textrm{ms}$.
The current amplitude in the top and bottom panel is $I_0=7.33\mu\textrm{A/cm}^2$,
and $7.303\mu\textrm{A/cm}^2$, respectively.
}
\label{fig15}
\end{figure}

An illustration of slow firing rates near the resonance is shown
in Fig. \ref{fig15}.
It is quite surprising that this aspect of the HH neuron dynamics
was not known until recently \cite{Borkowski2011}.
We can expect similar slowing down in many other models of resonant neurons as well,
although the exact details of this behavior will depend on the relative magnitude
of time scales of each model.

\begin{figure}[ht]
\center
\includegraphics[width=0.45\textwidth]{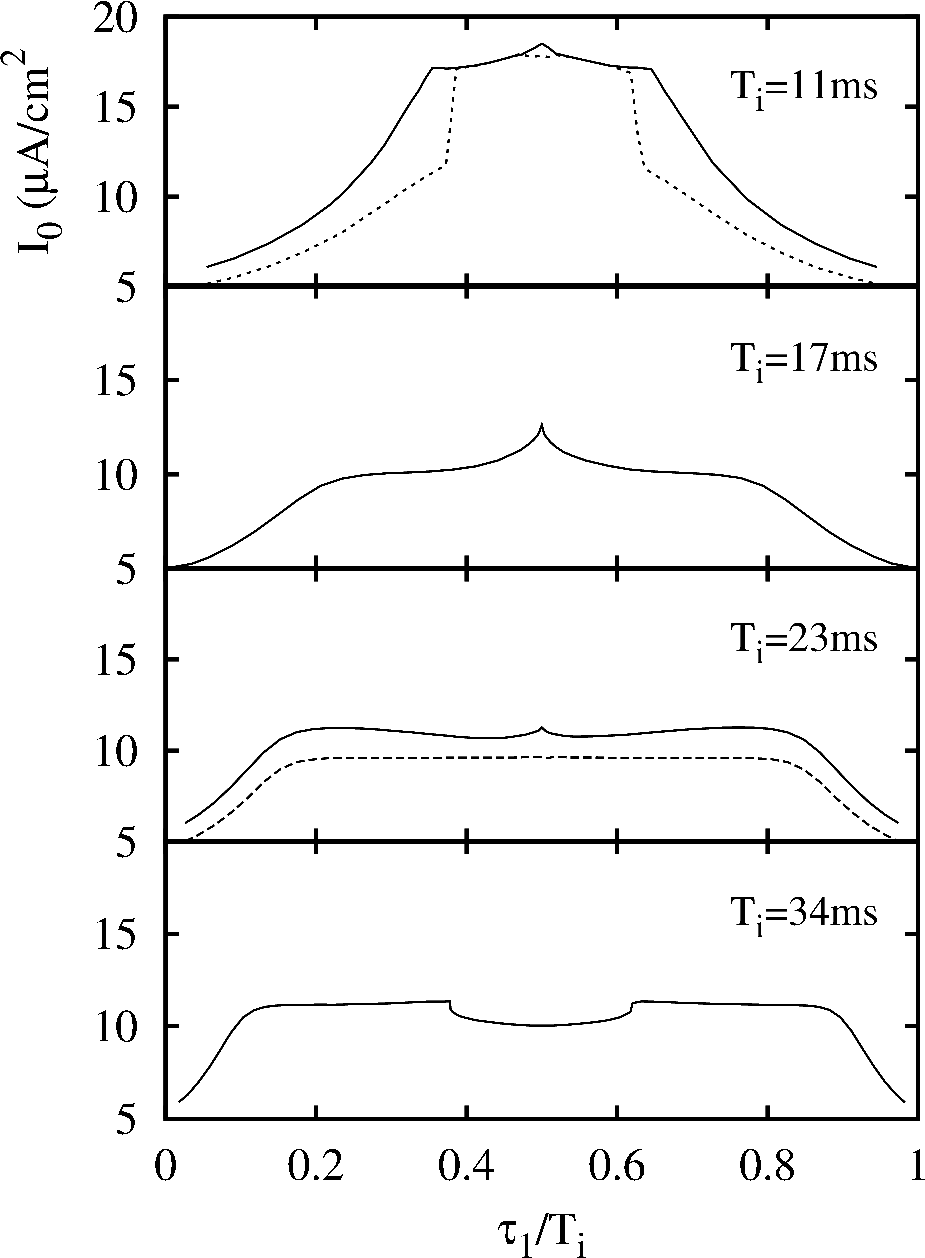}
\caption{Excitation edge for a double pulse as a function
of separation between the leading
and the trailing phase. The inter-phase gap is scaled in units of the stimulation
period $T_i$. The diagrams from top to bottom
are drawn for $T_i=11, 17, 23$, and $34\textrm{ms}$, respectively.
}
\label{fig16}
\end{figure}

The threshold for biphasic monopolar pulses is shown in Fig. \ref{fig16}.
For $T_i \le T_{res}$ the most regular input
signal with equally spaced pulses, $\tau_1=T_i/2$,
is also the least likely to elicit spiking.
The notation here is the same as in Fig. \ref{stimulus}.
Any deviation from the perfect regularity of the stimulus for $\tau_1=T_i/2$ and $T_i \le T_{res}$,
either by means of a deterministic or stochastic perturbation
leads to a decrease of threshold.

Comparing Fig. \ref{fig12} to Fig. \ref{fig2},
we can conclude that the topology of the response diagram
does not depend on the size of IPG for $\tau_1<5\textrm{ms}$.
The same is likely to be true for all
types of periodic short pulses irrespective
of details of their time dependence.
This is an illustration of the well known fact that the HH neuron
has a charge threshold property. When a relatively strong current
is delivered in a short time, the voltage
change is mainly determined by the capacitive current \cite{Hodgkin1946,Noble1966,Koch1999}.

\begin{figure}[ht]
\includegraphics[width=0.45\textwidth]{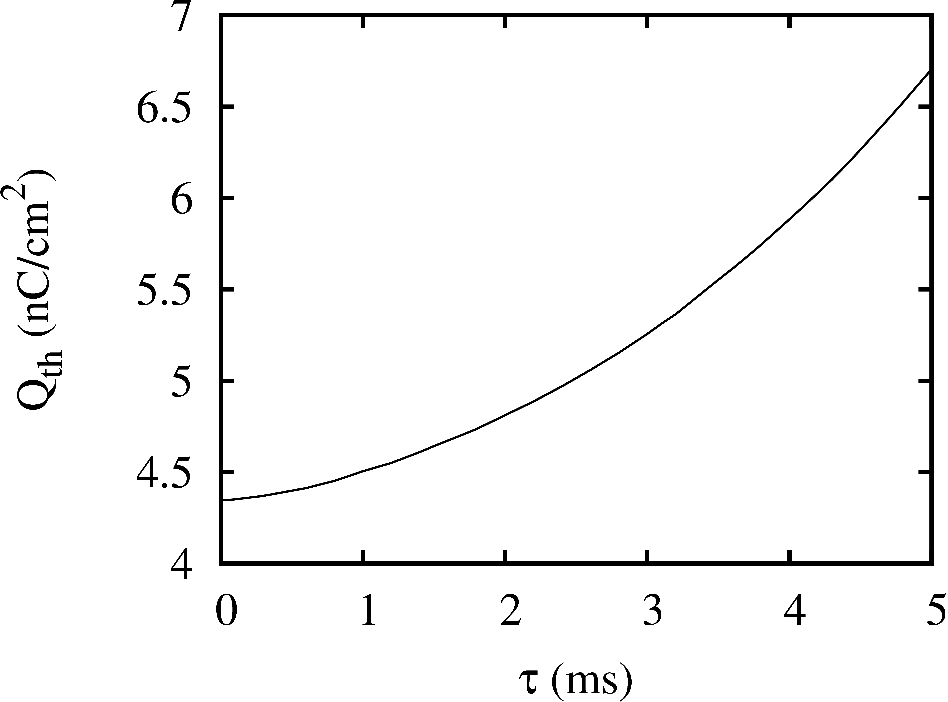}
\hskip 1cm
\includegraphics[width=0.45\textwidth]{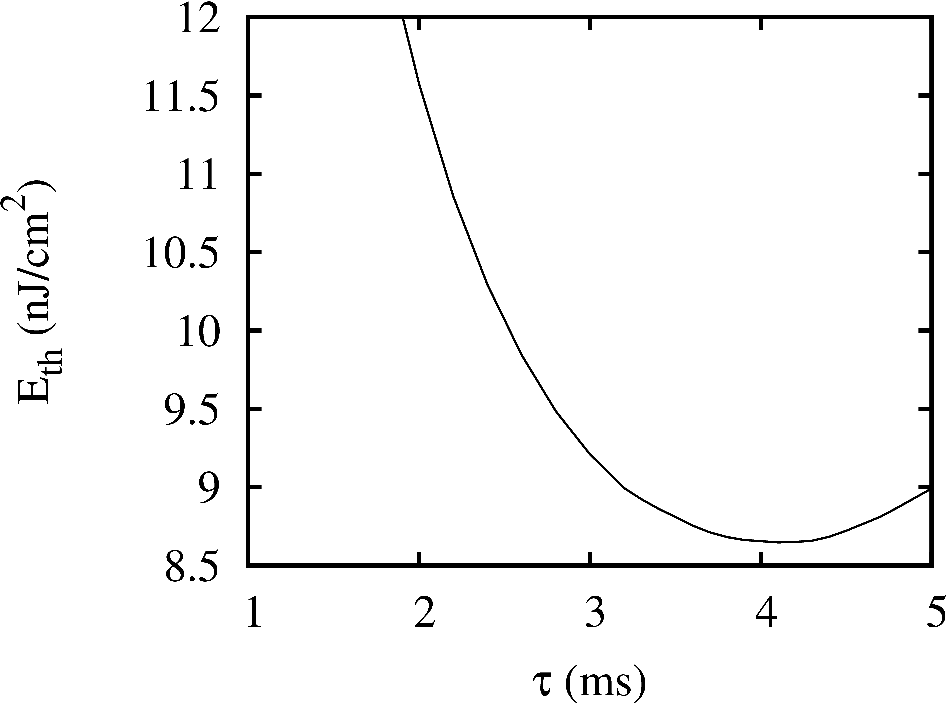}
\caption{Threshold charge (left panel) and threshold energy (right panel)
as a function of $\tau$
for a resonant stimulation, $T_i=17\textrm{ms}$, by a biphasic periodic sequence of pulses
with an optimal inter-phase gap, $\tau_1=5\textrm{ms}$.
The minimum charge occurs at $\tau \rightarrow 0$ which means
that the most effective pulses are those with the shortest possible width.
The pulse energy has a minimum near $\tau=4\textrm{ms}$.
}
\label{fig17}
\end{figure}

In neural electrical stimulation the main safety factors
are the amount of charge transferred in a single pulse, $Q_{th}$,
and energy per spike, $E_{th}$, needed to evoke spiking,

\begin{equation}
Q_{th} \sim \int_0^{\tau} I_{th}(t)dt
\end{equation}

\begin{equation}
E_{th} \sim \int_0^{\tau} I^2_{th}(t)dt,
\end{equation}
where $I_{th}$ is the threshold current amplitude.
Fig. \ref{fig17} shows the threshold charge
at $T_i=17\textrm{ms}$ as a function of the phase width $\tau$ for
$\tau_1=5\textrm{ms}$.
The charge-duration curve in Fig. \ref{fig17}
implies the use of short pulses in stimulation protocols
if the injected charge is to be minimized.
In practice, pulses lasting tens of microseconds approach
the minimum charge condition sufficiently well
and are often a reasonable solution in the design of neural prostheses.
During this relatively short time one may be
able to avoid Faradaic reactions that would occur at higher
levels of total charge with longer pulses.
Similar conclusion was reached in a study of Sahin and Tie \cite{Sahin2007}
who investigated effects of non-rectangular waveforms on threshold
both experimentally and theoretically within
a simple local model of a mammalian nerve \cite{Chiu1979,Sweeney1987}.
Their model is similar to the HH model except for the absence
of the potassium channel.
However if the main limiting factor is the energy
delivered per pulse, then finite $\tau$ is preferable,
see Fig. \ref{fig17}.
Qualitatively similar result was obtained
for different waveforms in Ref. \cite{Sahin2007}.

\section{Discussion}
The global bifurcation diagram of the HH neuron stimulated
by biphasic charge-balanced pulses is very similar to the diagram
obtained in the study of monophasic stimulation \cite{Borkowski2011}.
Charge-balancing has no effect on topology of the global
bifurcation diagram in the period-amplitude plane.
Fig. \ref{fig2} closely resembles results obtained
for the monophasic stimulation \cite{Borkowski2011}.
These properties are quite general, provided the width
of the current pulses is sufficiently small, which implies
$\tau_1<5\textrm{ms}$ for the classic HH parameter set,
and should hold for other, non-rectangular pulse shapes.

Studies of threshold behavior of resonant neurons may help design future
cochlear implants.
The excitation threshold is a very sensitive function
of IPG. When the anodic part of the pulse follows immediately the cathodic
part the magnitude of threshold current pulses at the main resonant frequency
is almost $I_0 \simeq 30 \mu\textrm{A/cm}^2$ (see Figs. \ref{fig11}
and \ref{fig2})
for the phase width of $\tau=0.6 \textrm{ms}$. When the IPG is increased
the threshold decreases, reaching $I_0 \simeq 7.3 \mu\textrm{A/cm}^2$
(see Fig. \ref{fig12})
for the near-optimal inter-phase separation of $\tau_1 = 5 \textrm{ms}$.
In the monophasic stimulation by current pulses
of the same phase width the threshold at $T_{res}$ is approximately
$I_0 \simeq 10 \mu\textrm{A/cm}^2$ \cite{Borkowski2011}.
It is interesting to note that in the monophasic case
the perfectly periodic stimulus has a higher threshold
than a signal with unequal intervals between subsequent pulses,
see Fig. \ref{fig16}.
Results in Figs. \ref{fig11} and \ref{fig17}
indicate that IPG giving the lowest firing
threshold of the HH membrane is approximately $4-5\textrm{ms}$.
The IPG required to minimize threshold is approximately the same as the IPG
which minimizes the energy delivered per pulse.
The decrease of threshold associated with the increase
of IPG is consistent with experimental measurements \cite{Shepherd1999}.
Similar result was obtained for biphasic pulses with
long IPG in a phenomenological
model of electrically stimulated human ANF \cite{Macherey2007}.
Our results are also consistent with the work
of Carlyon et al. \cite{Carlyon2005} who studied human behavioral thresholds
for trains of biphasic pulses applied to a single channel
of cochlear implants as a function of IPG.
The experimental threshold decreased for IPGs up to several milliseconds
when the phases of the pulse were of opposite polarity.

The firing threshold of the HH model
is a nonmonotonic function of the stimulation period with the minimum
at the main resonance.
A local minimum exists also at very high stimulation rate.
At the resonant stimulation period and its multiples the solution is always unique
and the firing rate is a continuous square-root
function of the pulse amplitude $I_0$, $f_o/f_i \simeq 2(I_0/I_{th}-1)^{1/2}$,
where $I_{th}$ is the threshold.
Thus $f_o$ can be arbitrarily low.
The simplest way to obtain desired interspike
separation $T_o$ in a resonant neuron of the HH type
is to stimulate it with current pulses of amplitude
\begin{equation}
\label{I0}
I_o \simeq I_{th} (1+\frac{T_{res}^2}{2T_o^2}) .
\end{equation}
Naturally, it should be kept in mind that a resonant neuron
need not have the dynamics of the HH model \cite{Borkowski2012}
and the square root dependence of the firing rate on current
amplitude is not always available.

The optimal stimulation in clinical applications may include Gaussian noise
superposed on the input signal \cite{Matsuoka2000a}. Small amounts of noise
have been shown to improve detection of faint signals, improve
temporal resolution and regularize strongly irregular firing
in regimes of dynamic instabilities.
Noise eliminates bistability at nonresonant frequencies \cite{Borkowski2011}.
However noise also broadens the local minima
of excitation threshold which is not desirable
in tasks such as selective stimulation of nerve fibers.

The recently discovered odd-all multimodal transition \cite{Borkowski2009,Borkowski2010},
related to the the vanishing of even response modes and accompanied
by a deep local minimum of the firing rate, occurs also
for charge-balanced stimuli.
It is located between the 2:1 and 3:1 locked-in states and is most pronounced
near excitation threshold.
This phenomenon is a manifestation of the internal neuron dynamics.
It is not related to a functional form of the stimulus.
The change of the current pulse waveform shifts the excitation threshold
but does not affect the topology of the global bifurcation diagram.
Also the frequency of the multimodal transition is invariant to the changes
of the pulse shape.

O'Gorman et al. proposed that firing irregularity of ANF
at high stimulation rates is due to a dynamic instability \cite{OGorman2009,OGorman2010}.
It is characterized by positive values of the Lyapunov exponent and explains both the sensitivity
to small changes of the stimulus and the lack of synchronization
of response of different ANFs to external drive \cite{OGorman2009}.
Although O'Gorman et al. analyzed the instability from a different
viewpoint its dynamical mechanism in the FHN model is the same
as in the HH model.
The MMT \cite{Borkowski2009,Borkowski2010}
and a dynamic instability \cite{OGorman2009,OGorman2010}
associated with it
are key properties of many resonant neurons.
MMT implies threshold bistability at nonresonant stimulation frequencies
above the natural frequency of the neuron \cite{Borkowski2012}.
The converse is not always true. Bistability at the threshold does not
always imply the existence of MMT.
As our recent study showed, neuron dynamics may be divided
into four classes \cite{Borkowski2012}. Only one of them displays MMT.
Thus it is important to correlate the threshold studies with the search
for the signatures of MMT in the ISI histograms.

This phenomenon may be relevant also in explaining the desynchronizing
effect of high frequency stimulation on networks of oscillating neurons.
Llinas et al. \cite{Llinas1999} claimed that
many neurological disorders are caused by 
pathological resonant interaction between cortical and sub-cortical structures.
Subsequent studies of network dynamics in Parkinson disease
found enhanced neural synchrony in the beta band (13 - 30 Hz) \cite{Hutchison2004}.
In fact, McIntyre and Hahn \cite{McIntyre2010} point out that
the strongest therapeutic effect in DBS treatment is obtained
at frequencies of order $100\textrm{Hz}$ and higher.
They also propose that the optimal DBS frequencies
are resonant frequencies of the cortico-basalganglia-thalamo-cortical
loop or loops within this network. In our work the dominant high-frequency
effects are of \textit{anti-resonant} character.
When some neurons in the loop
are forced to fire chaotically by high-frequency stimulation,
they are prevented from participating in a coordinated
activity of a neural ensemble at their natural frequencies.

The model considered here assumes intracellular current
injection. Since clinical neurostimulation devices
deliver current in the extracellular space, 
let us briefly comment how the presence of tissue
affects the problem. Most bioelectric field models 
assume a purely resistive tissue
and ignore capacitive, inductive and wave propagation effects.
This approximation may not always be appropriate for stimulation
by high-frequency pulses.
In a realistic calculation both electrode
and bulk tissue capacitance should be included \cite{Butson2005}.
Bossetti et al.\cite{Bossetti2008} studied the problem
and concluded that wave propagation
and inductive effects can usually be neglected.
However the quasi-static approximation is valid only for a specific
range of tissue dielectric properties.
The analysis of Tracey and Williams \cite{Tracey2011}
showed that attenuation due to frequency-independent capacitance
increases firing thresholds and the dispersion caused by the
frequency-dependent effects may have an opposite influence,
leading to lower threshold currents.
Foutz and McIntyre pointed out that optimal pulse width
depends on electrode geometry \cite{Foutz2010}.
In the case of extracellular stimulation
we expect the global bifurcation diagram, Figs. \ref{fig2}
and \ref{fig12},
to have the same overall form, although details such as precise location
of the threshold would be affected. MMT would occur
at the same frequencies since it is a reflection of the internal
dynamics of the neuron.

\section{Summary of results}

We would like to conclude the article by stating its main results:

\begin{itemize}
\item{
Stimulation by periodic biphasic pulses gives the same global bifurcation
diagram in the period-amplitude plane as stimulation by monophasic pulses.}

\item{There is a multimodal transition between the mode-locked states
2:1 and 3:1. This is a line of critical points $T_{mm},I_{mm}$ such that
below the transition, where $T_i < T_{mm}$, only odd response modes exist.
Even modes appear for $T_{mm} < T_i$.}

\item{
We discovered a new phenomenon: the vanishing
of the modes $2,5,8,\ldots$ in the regime
of irregular firing between the 3:1 and 4:1 states.
}

\item{
The maximum threshold is obtained near $T_i \simeq 5.5\textrm{ms}$,
which is of order $T_{res}/3$,
between the entrained states 3:1 and 4:1.}

\item{
The minimum threshold at the tip of the main resonance
is obtained for the inter-phase gap of about $5\textrm{ms}$.
This value of IPG is also associated with the minimum energy needed
to elicit the response.}

\item{
The location of the minimum threshold depends on the size
of IPG. For biphasic pulses with $\tau=\tau_1=0.6\textrm{ms}$
the tip of the resonance
is located at $T_i\simeq 17.5\textrm{ms}$. For increasing
IPG the resonance moves to higher values, reaching
$18.3\textrm{ms}$ for $\tau_1=5\textrm{ms}$.}

\item{
The firing frequency is a continuous, square root function of the pulse
amplitude at the resonance.}

\item{The resonant neuron of the HH type can be forced
to fire with arbitrarily low frequency $f_o$
by delivering current pulses of amplitude
$I_o \simeq I_{th} (1+f_o^2/2f_{res}^2)$
}

\item{
The antiresonant effects at high stimulation frequency, such as the multimodal transition,
may help explain the therapeutic mechanism of DBS.
}

\end{itemize}



\end{document}